\documentclass[a4paper,11pt]{article}
\usepackage[left=2.5cm,top=3cm,right=2.5cm,bottom=3cm,bindingoffset=0.0cm]{geometry}

\usepackage[utf8]{inputenc}  % Allow UTF-8 encoded characters
\usepackage[T1]{fontenc}     % Better accented characters
\usepackage{amsmath, amssymb}
\usepackage{graphicx}        % Include figures
\usepackage{hyperref}        % Hyperlinks in PDF
\usepackage{booktabs} % Enhances table formatting with better rules (lines)
\usepackage{siunitx}
\usepackage{subcaption}
\usepackage[backend=biber,
            style=numeric,
            datamodel=standard,
            sorting=none,
            doi=true,
            url=false,
            isbn=false]{biblatex}
\title{Impacts of internal heating on temperature distribution in channels}

\usepackage{authblk}
\author[1]{Lubomír Bureš}
\author[1]{Per Nilsson}
\affil[1]{Saltfoss Energy ApS, Titangade 11, 2200 Copenhagen, Denmark \vspace{1ex}}
\affil[ ]{ \texttt {\{lubomir.bures,per.nilsson\}@saltfoss.com}}
\date{July 11, 2025}

\begin{document}
\maketitle

% Abstract
\begin{abstract}
  \noindent In molten-salt-fuelled reactor systems, the fluid may experience substantial volumetric heat generation in addition to heat removal from surrounding structures. To quantify these effects, we investigate developed channel flow with internal heating using a systematic multi-scale approach comprising Direct Numerical Simulation (DNS), Large Eddy Simulation (LES), and a semi-analytical solver (SAS). First, DNS and LES are compared in a turbulent parallel-plate configuration at different Prandtl and Reynolds numbers, demonstrating excellent agreement in flow and thermal fields, with the SAS method showing acceptable accuracy. Building on this benchmarking, the SAS tool is then employed to explore a broad parameter space, offering insights into how internal heat deposition modifies the temperature distribution across Reynolds and Prandtl numbers. Comparisons are also drawn against the canonical wall-heating scenario, revealing that volumetric heating often remains a secondary effect in turbulent regimes but can become more pronounced at lower Reynolds numbers, higher Prandtl numbers, or when nearly all heat is deposited in the fluid. These findings establish guidelines for reduced-order modeling in liquid-fuel reactor analyses and highlight conditions under which internal heating warrants particular attention. The paper concludes by outlining ongoing and future research directions, including refinements for variable fluid properties and complex geometry extensions.

  \par\medskip\noindent
  \textbf{Keywords:} molten-salt reactor; internal heat generation; direct numerical simulation; large eddy simulation; heat transfer
\end{abstract}

% Main content
\section{Introduction}
Analysis of fluid temperature distribution is one of the primary applications of thermal fluid dynamics. This becomes especially relevant in industrial settings, in which fluids are used to remove heat from solid structures, ensuring their temperatures stay below limits imposed by material properties. In conventional nuclear applications, the chief object of study is the assessment of heat-transfer performance of the coolant in the reactor core under various operational and postulated accidental regimes. To this end, a myriad of studies have been performed up to the present day resulting in the development of trusted heat-transfer correlations for the Nusselt number $\textit{Nu}$ (-), 
\begin{equation}
	\textit{Nu} = \frac{\alpha D}{\lambda},
\end{equation}
where $\lambda$ (W/m/K) is the coolant thermal conductivity, $D$ (m) the characteristic dimension (typically the hydraulic diameter $D_h$), and $\alpha$ (W/m$^2$/K) the wall-bulk heat-transfer coefficient appearing in the closure relation for applied wall heat flux $j$ (W/m$^2$) as given by the Newton's law of cooling:
\begin{equation}
	j = \alpha (T_w - T_b),
\end{equation}
with $T_w$ (K) being the wall temperature and $T_b$ the coolant bulk temperature. This can be formally re-arranged as:
\begin{equation}
	T_w = T_b + \frac{j}{\alpha} = T_b + \frac{j D}{\lambda\textit{Nu}},
	\label{eq:tw_jw}
\end{equation}
indicating our interest in the wall temperature. Correlations for $\textit{Nu}$ (see e.g.\ \cite{Meyer2019} for a review for circular pipes) are underpinned by extensive experimental, theoretical, and computational studies. On the computational side, the well-established multi-scale approach from Direct Numerical Simulation (DNS) through Large-Eddy Simulation (LES) and Reynolds-Averaged Navier Stokes (RANS) Computational Fluid Dynamics (CFD) to system thermal-hydraulic modelling allows for the development and application of practical reduced-order models even in cases where robust experimental data are not available.

In the special case of liquid-fuelled nuclear reactors, nowadays primarily represented by molten-salt reactors, the applicability of conventional heat-transfer correlations is limited. This is partially due to the use of non-conventional liquids, such as molten fluoride salts, but especially because of the fact that the heat generated in the reactor core is deposited primarily in the liquid fuel/coolant. This means that Eq.\ \ref{eq:tw_jw} should be re-cast in the form:
\begin{equation}
	T_w = T_b + \frac{j D}{\lambda\textit{Nu}_j} + \frac{q D^2}{\lambda \textit{Nu}_q}.
	\label{eq:tw}
\end{equation}
Here, $q$ (W/m$^3$) is the fluid volumetric heat source, $\textit{Nu}_j$ the conventional Nusselt number introduced above and $\textit{Nu}_q$ the Nusselt number due to internal heating, signifying that even in the absence of inward heat flux, the wall temperature will be elevated due to bulk heat generation. 

The effect of internal heating on fluid temperature distribution has been studied before, the seminal work of Poppendiek and Palmer established analytical solutions for developed laminar flow in circular-pipe \cite{Poppendiek1952} and parallel-plate \cite{Poppendiek1954} geometries and proposed a semi-analytical approach to derive a solution for turbulent flow configurations as well. The key insight to reduce the dimensionality of this problem already present in \cite{Poppendiek1952} is that, in case that due to the linearity of the governing equations in the case of fluid material properties assumed independent of temperature, the temperature field can be decomposed in two; one corresponding to a flow system with non-zero wall heat flux and no internal heat generation and the other to one with no wall heat flux and non-zero internal heat generation. For example, the overall parallel-plate analytical solution for temperature as a function of the distance from the median plane $r$ is ($\delta$ being the channel half-width) \cite{Poppendiek1954}:
\begin{equation}
	\frac{T(r) - T_b}{\delta/\lambda} = \frac{3q\delta+17j}{35} + \frac{q\delta+3j}{4}\bigg[\bigg(\frac{r}{\delta}\bigg)^2 - 1\bigg] - \frac{q\delta+j}{8}\bigg[\bigg(\frac{r}{\delta}\bigg)^4 - 1\bigg].
	\label{eq:poppendiek}
\end{equation}
For $r=\delta$ it is obtained:
\begin{equation}
	T_w  = T_b + \frac{17j \delta}{35\lambda} + \frac{3q\delta^2}{35\lambda} = T_b + \frac{17j D_h}{140\lambda} + \frac{3qD_h^2}{560\lambda},
	\label{eq:tw_lam}
\end{equation}
and the canonical Nusselt numbers for infinite parallel-plate geometries with Neumann wall boundary condition for temperature can be recovered as:
\begin{align}
	\textit{Nu}_j &= 140/17 \approx 8.235, \label{eq:nuj_laminar}\\
	\textit{Nu}_q &= 560/3 \approx 186.7. \label{eq:nuq_laminar}
\end{align}

In recent years, the focus has been on analysing the effect of internal heating under turbulent conditions, especially in circular pipes. To this end, a semi-analytical approach for reduced-order modelling of the turbulent Graetz problem with a volumetric heat source was developed and successfully used to derive heat-transfer correlations in \cite{DiMarcello2010,Fiorina2013,Fiorina2014} with comparisons made against RANS in \cite{Luzzi2010} indicating good agreement. Further RANS analyses and method development were also performed in \cite{Fiorina2019} and \cite{Cammi2024}. 

In spite of this significant effort, to the best of the authors' knowledge, no systematic multi-scale model development and benchmarking have been performed so far for turbulent flow with internal heat generation. This limits the perceived degree of reliability of existing numerical results. For example, RANS, the approach of Poppendiek and Palmer as well as the semi-analytical solver (SAS) of \cite{DiMarcello2010} all rely on reduced-order modelling, especially in the near-wall region. Due to the theoretical similarity of the employed closure laws, the value of successful benchmarking of these approaches against one another is potentially diminished. The focus of our present work is to remedy this issue. Using high-fidelity DNS as a foundation, we can progressively reduce the computational requirements by using first LES and then SAS, quantifying the error of these approximations for the simulation of internally-heated flows, and supplying confidence in the overall approach.

In this paper, we start by introducing our computational methods in Section \ref{sec:methods}. Then, in Section \ref{sec:benchmarking}, we compare the results of our models against one another, quantifying the error of the reduced-order approaches and in Section \ref{sec:rom} we comment on general trends of internal heat generation effects for different values of governing parameters. The paper is concluded in Section \ref{sec:conc}, where the directions for future work are also indicated.

\section{Methods}
\label{sec:methods}
\begin{figure}
	\centering
	\includegraphics[width=0.7\linewidth]{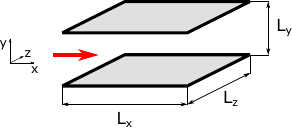}
	\caption{Geometry of the problem under consideration. The domain is periodic in the $x$ and $z$ directions and wall-bounded in the $y$ direction. Constant mean velocity $u_b$ was considered in the $x$ direction and constant volumetric heat generation $q$ was assumed in the whole domain. The walls are adiabatic.}
	\label{fig:drawing}
\end{figure}

In this section, the computational methods used are described. In all cases, the same conceptual problem was simulated, i.e.\ bounded periodic flow between two infinite parallel plates with internal heating as shown in Fig.\ \ref{fig:drawing}. However, different approaches were taken for the individual computational methods.

\subsection{Direct numerical simulation}
For DNS, the high-order spectral element code Nek5000 \cite{Fischer2007} was chosen. For this study, Nek5000 was used to solve the incompressible Navier-Stokes equations coupled with the energy transport equation for constant-property fluids and the equations were cast in a non-dimensional form. Furthermore, instead of absolute temperature $T$, the temperature deviation from the bulk mean $\theta = T - T_b$ was considered, eliminating the issue of unbounded temperature increase due to heat generation imbalance, and constant mean velocity in the $x$ direction was prescribed. This followed the approach described in \cite{Kawamura2000,Prasad2016} and for the problem under consideration the governing equations took the form:
\begin{align}
  \frac{\partial \tilde{u}_i}{\partial \tilde{x}_i} &= 0, \\
\frac{\partial \tilde{u}_i}{\partial \tilde{t}} + \tilde{u}_j\frac{\partial \tilde{u}_i}{\partial \tilde{x}_j} &= -\frac{\partial \tilde{p}}{\partial \tilde{x}_i} + \frac{1}{\textit{Re}_\delta}\frac{\partial^2 \tilde{u}_i}{\partial\tilde{x}_j^2}, \\
\frac{\partial \tilde{\theta}}{\partial\tilde{t}} + \tilde{u}_j\frac{\partial \tilde{\theta}}{\partial \tilde{x}_j}  &= \frac{1}{\textit{Re}_\delta \textit{Pr}}\frac{\partial^2 \tilde{\theta}}{\partial \tilde{x}_j^2} - \tilde{u}_x + G.\\
  \frac{1}{Re_\delta Pr} \frac{\partial \tilde{\theta}}{\partial \tilde{y}}&\bigg|_{w, \text{ inwards}} = 1 - G.
\end{align}
Here, $\sim$ indicates dimensionless quantities with the characteristic length $x_c = \delta = L_y / 2$, the characteristic velocity $u_c = u_b$ (the mean flow velocity), and the characteristic temperature:
\begin{equation}
	\theta_c = \frac{q \delta + j}{\rho c_p u_b},
	\label{eq:thetac}
\end{equation}
with $\rho$ being the fluid density (kg/m$^3$) and $c_p$ the specific heat capacity (J/kg/K). In the problem under consideration, $j=0$. Furthermore, in the governing equations, $\textit{Pr}$ is the Prandtl number and the other two dimensionless numbers are defined as:
\begin{align}
	\textit{Re}_\delta &= \frac{u_b \delta}{\nu}, \label{eq:rem} \\
	G &= \frac{q\delta}{q\delta + j}, \label{eq:g}
\end{align}
i.e.\ $G=1$ here. In Eq.\ \ref{eq:rem}, $\nu$ is the kinematic viscosity (m$^2$/s).

In this work, we performed DNS at friction Reynolds number $\textit{Re}_\tau=180$, where:
\begin{equation}
	\textit{Re}_\tau = \frac{u_\tau\delta}{\nu},
\end{equation}
$u_\tau$ being the friction velocity. This corresponds to $\textit{Re}_\delta \approxeq 2813$, which was prescribed in the simulation. We simulated three different Prandtl numbers: 1, 7, and 15. For this Reynolds number and the two lower Prandtl numbers, high-quality reference DNS data are available for wall-bounded flows without and with heat transfer \cite{Moser2007,Seki2008}. These references were used in preliminary simulations for $G=0$ (i.e.\ only wall heat transfer and no internal heat generation) to give confidence in the sufficient convergence of the solution for chosen mesh resolution and simulation duration. The simulation setup was based on the publicly available turbulent Nek5000 case \cite{turbChannel} and the analysis done in \cite{Prasad2016}. The mesh was generated using the built-in Nek5000 \texttt{genbox} generator. In $x$ and $z$ directions, a uniform mesh was used with 16 elements, $\tilde{L}_x = 2\pi$ and $\tilde{L}_z = \pi$. In the $y$ direction, $\tilde{L}_y = 2$ by definition and 20 mesh elements were used with mesh grading defined by the coordinate transformation:
\begin{equation}
	\tilde{y} = \tanh(2.4\xi)/\tanh(2.4),\ \xi \in [-1,1].
\end{equation}
Eight (8) quadrature points were used per element per direction. This setting corresponded to the first near-wall value of $y_{+,1} \approx 0.146$, defined as:
\begin{equation}
	y_+ = \frac{u_\tau y_w}{\nu},
	\label{eq:yplus}
\end{equation}
with $y_w$ being the distance to the nearest wall. Regarding time stepping, a variable time step according to the Courant-Friedrichs-Lewy condition $\textit{CFL}\leq0.5$ was used.

\subsection{Large eddy simulation}
For LES, the OpenFOAM v12 \cite{Greenshields2024} CFD software was chosen, employing the finite-volume method \texttt{incompre\-ssibleFluid} solver module for the incompressible Navier-Stokes equations and the \texttt{scalarTran\-sport} solver module for thermal energy density $e = \rho c_p T$. The governing equations were cast in their dimensional form with the Navier-Stokes equation formulation based on kinematic pressure and constant kinematic viscosity $\nu = 2 \cdot 10^{-5}$ m$^2$/s. The solution variables were implicitly filtered by the mesh grid with its lower limit on resolution. The sub-grid scales were modelled using the WALE model \cite{Nicoud1999, Mukha2015, Komen2020} with the default $C_w = 0.325$ for velocity and using Reynolds analogy for sub-grid-scale thermal energy with turbulent Prandtl number $\textit{Pr}_t = 1$. The need for a wall model was precluded by resolving the flow using a grid with a sufficient number of cells in the viscous layer, i.e.\ the approach which could be called ``wall-resolved LES'' was used. The chosen solvers and schemes for pressure and momentum, as well as solution tolerances were the same as the default settings in the OpenFOAM \texttt{channel395} tutorial case. They were second-order accurate in space and time and the PIMPLE implementation of the PISO algorithm was used.

The problem considered here was set-up as periodic by means of a stream-wise ($x$) momentum gradient using the built-in \texttt{meanVelocityForce} constraint. This effectively added a source term to the governing equations, which was iteratively tuned to a target bulk flow velocity. Zero wall heat flux was used as a boundary condition for energy with a step change in thermal energy $\Delta e$ applied between the inlet and outlet in the $x$ direction to compensate for internal heat generation. The simulation domain size was taken as $L_x \times Ly \times L_z = 6 \text{ m} \times 3 \text{ m} \times 2 \text{ m}$ and generated with \texttt{blockMesh}. Uniform discretization with 120 cells in the $x$ direction and 90 cells in the $z$ direction was used. In the $y$ direction, 100 cells were used with a logarithmic grading of 10 from the walls, i.e.\ the two cell layers at the channel median plane were ten times thicker than the layer at the walls. Overall, this gave a stretching ratio of less than 1.05, which was previously deemed satisfactory in \cite{Komen2020}. The wall-cell $y$ thickness is about $5\cdot 10^{-3}\cdot\delta$, which was considered sufficient in \cite{Mukha2015}. Regarding time stepping, a constant time step with $\textit{CFL}\leq1$ was used and the solution was iterated to statistical stationarity before gathering statistics. As a preliminary step, benchmarking against reference DNS data from \cite{Moser2007,Seki2008} and handbook formulas \cite{Reichardt1951,Kader1981} was conducted. Acceptable agreement was found overall; the performance was particularly good at lower Reynolds and Prandtl numbers. The sensitivity to the choice of sub-grid-scale Prandtl number was found to be rather low.

\begin{table}
\centering
\caption{LES case details.}
\label{tab:LEScases}
\begin{tabular}{lcccccc}
	\toprule
	$\textit{Re}_\tau$ (target) &  &180   &    &   & 395  & \\
	$\textit{Re}_\delta$ (set) &   & 2920  &    &   & 7170 & \\
	$\textit{Re}_\tau$ (result) &  &183   &    &   & 389  & \\
	$y_{+,1}$ (result) &   & 0.446 &    &   & 0.948 & \\
	
	\midrule
	$\Delta e$ (set) &   & 103 J/m$^3$ &    &   & 41.8 J/m$^3$ & \\
	
	\bottomrule
\end{tabular}
\end{table}

In this work, we performed LES at two friction Reynolds numbers $\textit{Re}_\tau=180$ and $\textit{Re}_\tau=395$, again with Prandtl numbers 1, 7, and 15, i.e.\ six (6) simulation cases in total. Since energy was treated as a passive scalar, it was only necessary to perform one distinct simulation for a given Reynolds number with multiple passive scalars transported with the flow at the same time. The characteristics of the simulated cases are given in Table \ref{tab:LEScases}. The bulk flow velocity $u_b$ was set based on the target friction Reynolds number using the Blasius friction factor formula. Therefore, some differences in the actual values of $\textit{Re}_\tau$ obtained in the simulation can be observed. Nevertheless, we will refer to the cases by their target  $\textit{Re}_\tau$ in the following text, since the impact of the minor variation of actual $\textit{Re}_\tau$ is considered to be low. The volumetric heat deposition was set to unity, $q = 1 \text{ W/m}^3$. As the $y_+$ of the centre of the cells adjacent to the walls was $\lesssim1$, the mesh was deemed sufficient for capturing the thermal field at $\textit{Pr}=1$. For higher Prandtl numbers, for which the thermal gradients are steeper than flow gradients, it could be argued that higher resolution would be required. This would, however, only be needed in regions where diffusive processes are significant in comparison with turbulent transport. For the investigated cases, this corresponded to the near-wall region. At the same time, the heat flux was set to zero at the wall, precluding the occurrence of significant temperature gradients. Therefore, the base resolution was deemed to be sufficient for thermal energy transport at all Prandtl numbers considered here.

\subsection{Semi-analytical solution}
The theoretical model presented in \cite{DiMarcello2010} was re-cast in the Cartesian coordinate system, implemented in Python, and solved. Even though the original model was developed to solve the Graetz problem for thermally developing flow, the stationary solution can be extracted as (given here only for the special case of zero wall heat flux and constant internal heat generation for simplicity):
\begin{equation}
	\tilde{\theta}(\tilde{r}) = \textit{Pr}\cdot \textit{Re}_\delta \cdot \bigg(\int_{0}^{1}\frac{h^2(\tilde{r})}{g(\tilde{r})}d\tilde{r} -\int_{\tilde{r}}^{1}\frac{h(\tilde{r}')}{g(\tilde{r}')}d\tilde{r}' + \int_{\tilde{r}}^{1}\frac{\tilde{r}'}{g(\tilde{r}')}d\tilde{r}' - \int_{0}^{1}\frac{h(\tilde{r})\tilde{r}}{g(\tilde{r})}d\tilde{r}\bigg).
	\label{eq:dimarcello}
\end{equation}
where:
\begin{align}
	h(\tilde{r}) &= \int_{0}^{\tilde{r}}\tilde{u}_x(\tilde{r}')d\tilde{r}',\\
	g(\tilde{r}) &= 1 + \textit{Pr}\cdot \textit{Re}_\delta\cdot  \frac{\tilde{\varepsilon}_m(\tilde{r})}{\textit{Pr}_t(\tilde{r})},
\end{align}
$\tilde{r}$ is the $y$ coordinate normalised by $\delta$ and taken to be zero at the channel median plane, $\tilde{\varepsilon}_m$ is the dimensionless eddy momentum diffusivity, and $\tilde{u}_x(\tilde{r})$, $\tilde{\varepsilon}_M(\tilde{r})$, $\textit{Pr}_t(\tilde{r})$ are found using the approach described in \cite{DiMarcello2010}. We note in passing that by setting:
\begin{align}
	\tilde{u}_x(\tilde{r}') &= \frac{3}{2}(1-\tilde{r}'^2),\\
	g(\tilde{r}) &= 1,
\end{align}
i.e.\ by converting the problem to a laminar flow case, the non-dimensionalised solution of Poppendiek and Palmer (Eq.\ \ref{eq:poppendiek}) is immediately recovered. Turning back to the problem under consideration, Eq.\ \ref{eq:dimarcello} and the associated equation for the velocity field were discretised using \num{1000} mesh points and the integrals were solved using the \texttt{numpy.trapezoid} method.

The SAS approach was benchmarked in this work against the aforementioned DNS and LES solutions. Note that for the $\textit{Re}_\tau = 180$ and $\textit{Re}_\tau = 395$ cases, the corresponding $Re_\delta$ values were 2748 and 6830, respectively  (cf.\ Table \ref{tab:LEScases}). The SAS was subsequently used for reduced-order parametric exploration, see Section \ref{sec:rom}.

\section{Benchmarking}
\label{sec:benchmarking}

\begin{figure}
	\centering
	% Row 1
	\begin{subfigure}{0.46\textwidth}
		\centering
		\includegraphics[width=\linewidth]{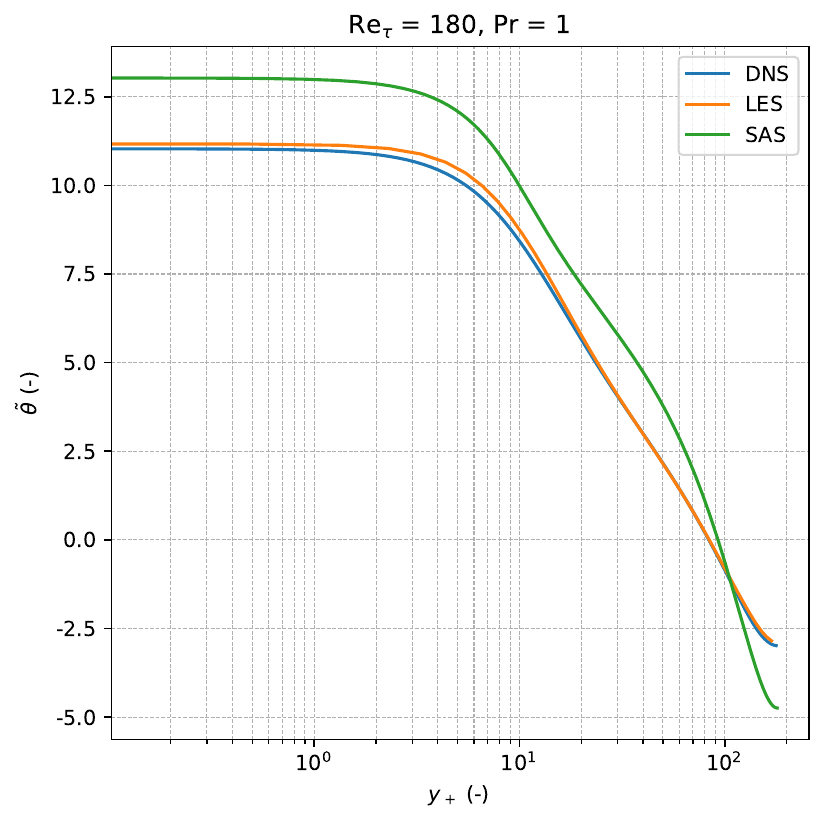}
	\end{subfigure}
	\begin{subfigure}{0.46\textwidth}
		\centering
		\includegraphics[width=\linewidth]{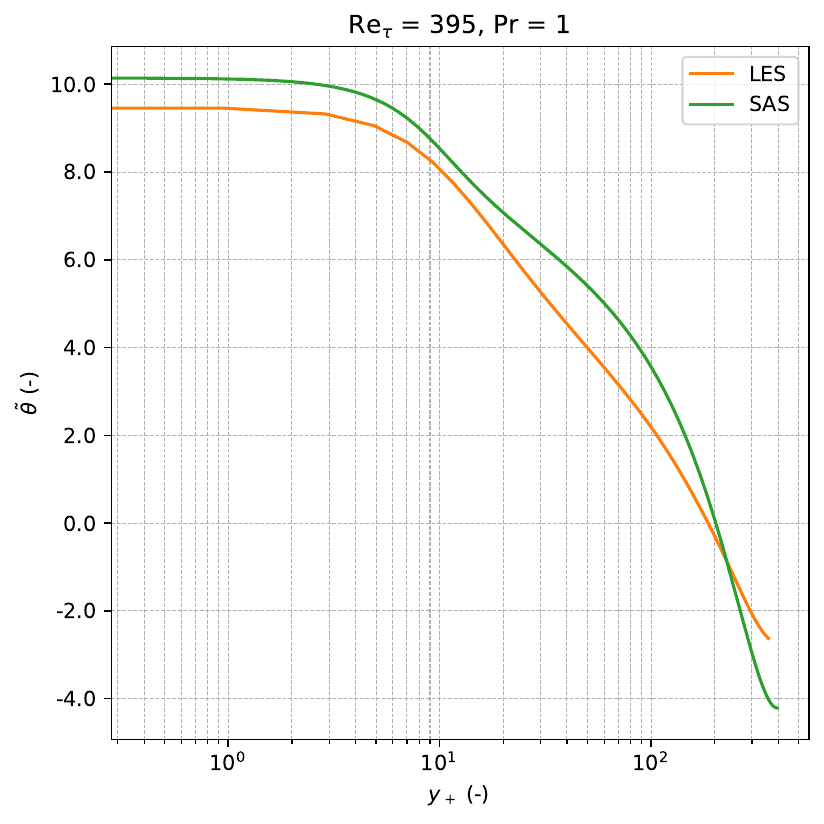}
	\end{subfigure}
	
	% Row 2
	\begin{subfigure}{0.46\textwidth}
		\centering
		\includegraphics[width=\linewidth]{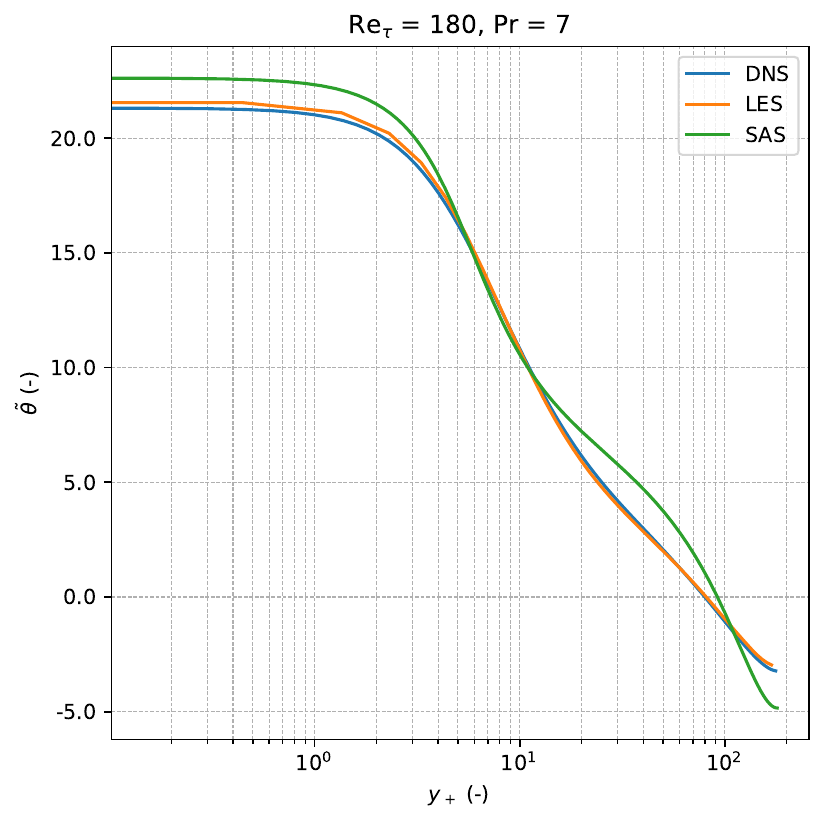}
	\end{subfigure}
	\begin{subfigure}{0.46\textwidth}
		\centering
		\includegraphics[width=\linewidth]{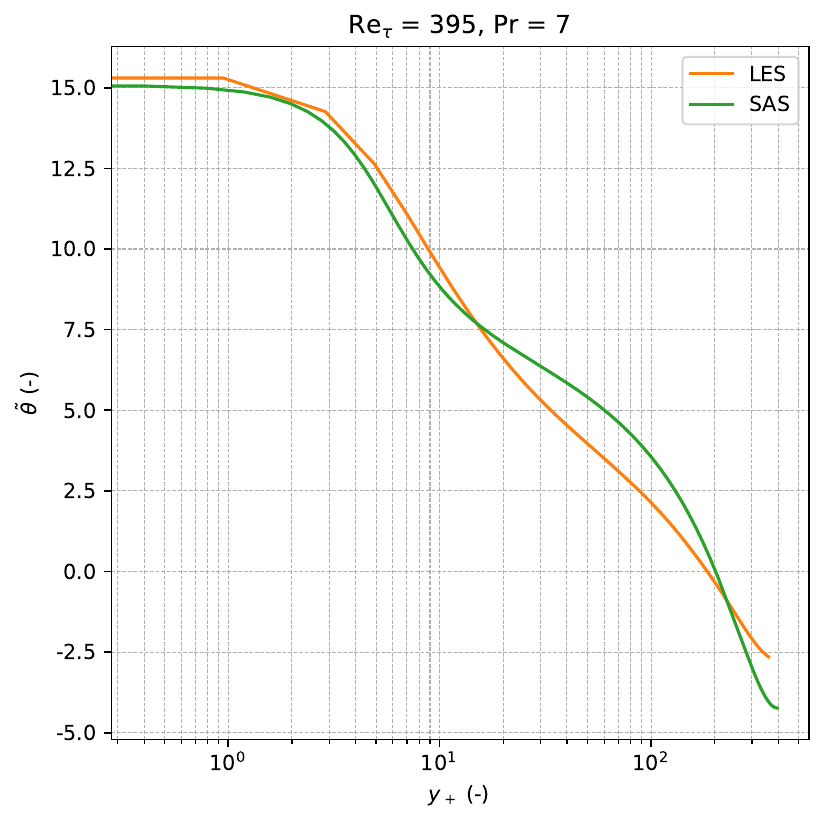}
	\end{subfigure}
	
	% Row 3
	\begin{subfigure}{0.46\textwidth}
		\centering
		\includegraphics[width=\linewidth]{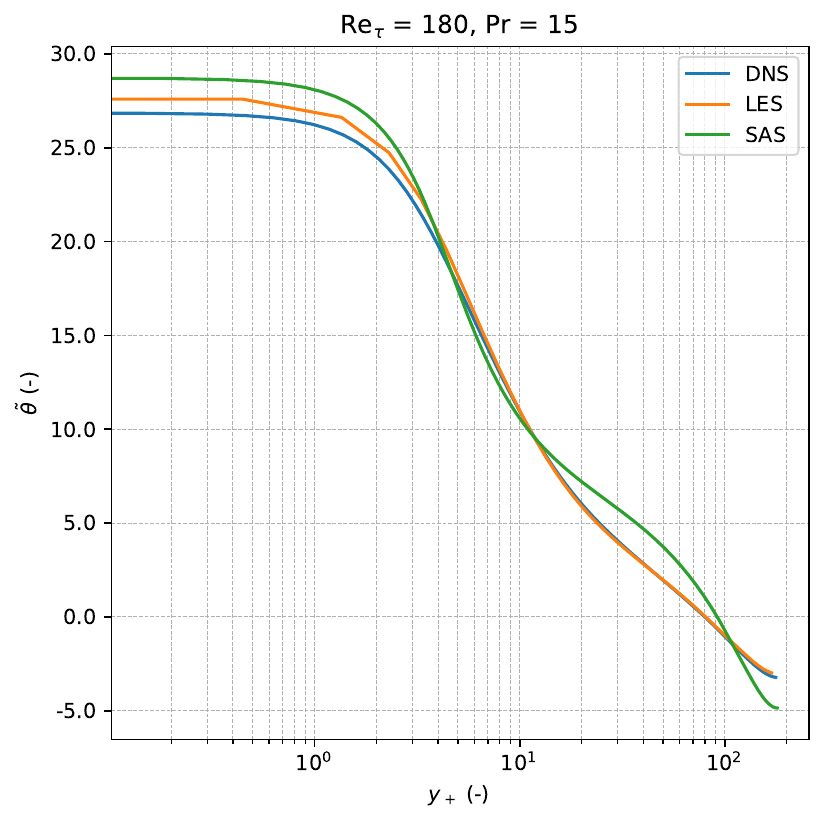}
	\end{subfigure}
	\begin{subfigure}{0.46\textwidth}
		\centering
		\includegraphics[width=\linewidth]{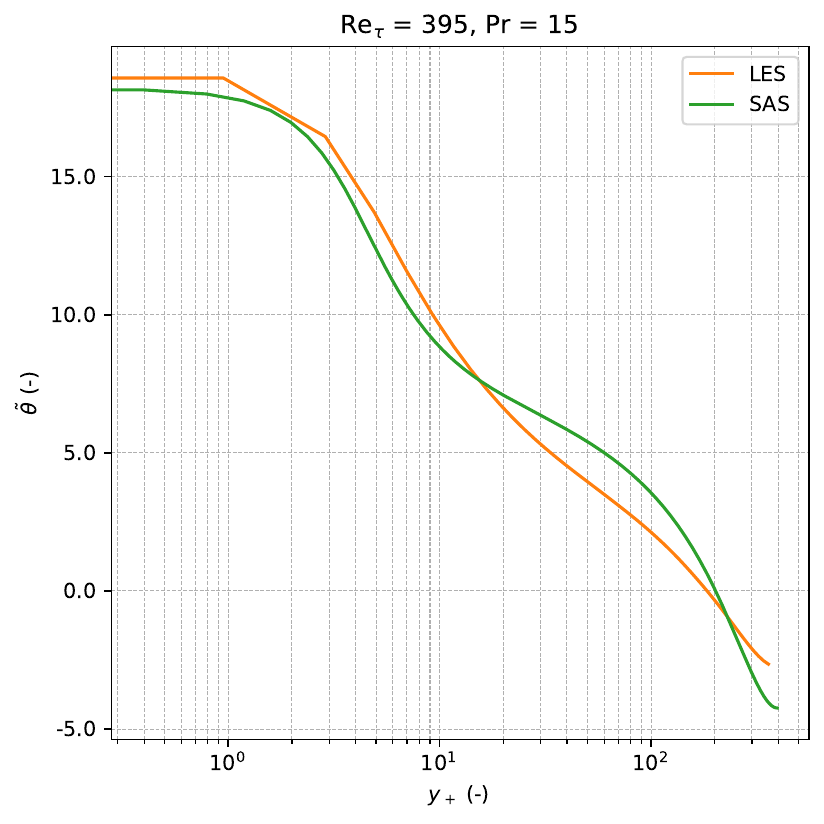}
	\end{subfigure}
	
	\caption{Results of the benchmarking exercise for considered values of $\textit{Re}_\tau$ and $\textit{Pr}$. In all plots, values of $\tilde{\theta}$, Eq.\ \ref{eq:thetatilde}, are given as functions of $y_+$, Eq.\ \ref{eq:yplus}.}
	\label{fig:benchmarking}
\end{figure}

In this section and the following one, we base our analysis on the non-dimensional temperature $\tilde{\theta}$, normalised by subtracting mean temperature rise and dividing by the characteristic temperature (Eq. \ref{eq:thetac} with $j=0$):
\begin{equation}
	\tilde{\theta} = \frac{T - T_b}{\theta_c} = \frac{T - T_b}{q\delta / (\rho c_p u_b)} = \textit{Pr} \cdot Re_\delta \frac{T - T_b}{q\delta^2 / \lambda}.
	\label{eq:thetatilde}
\end{equation}
Therefore, all the following discussions are primarily related to the \textit{shape} of the temperature field. For example, when it is stated that internal heating has limited impact, this should be understood in the context of the variations of temperature with respect to the transversal coordinate $y$. In other words, we do not concern ourselves with the effect of internal heating on the mean temperature rise in the channel.

Our first task  was to use the computational methods presented above for the comparison of calculated $\tilde{\theta}$ profiles. In Fig.\ \ref{fig:benchmarking}, the results for the three approaches, i.e.\ DNS with Nek5000, LES with OpenFOAM, and SAS, are plotted as functions of $y_+$ (Eq.\ \ref{eq:yplus}). Only one half of the channel is shown. For SAS, the values given in the figure were computed directly via Eq.\ \ref{eq:dimarcello}. For LES and DNS, statistical averaging was used. Note that DNS results are only available for $\textit{Re}_\tau = 180$. The plotted comparisons are complemented by quantitative data for $\tilde{\theta}(0)=\tilde{\theta}(y_+=0)$ in Table \ref{tab:benchmarking}.

\begin{table}
	\caption{Values of $\tilde{\theta}(0)$ and percentage differences between cases for values of $\textit{Re}_\tau$ and $\textit{Pr}$ considered in the benchmarking exercise.}
	\centering
	\begin{tabular}{c|cccccc}
		\toprule
		$\textit{Re}_\tau$ & \multicolumn{3}{c}{180} & \multicolumn{3}{c}{395} \\
		\midrule 
		$\textit{Pr}$ & 1 & 7 & 15 & 1 & 17 & 15 \\ 
		\midrule
	DNS & 11.0 & 21.3 & 26.8 & - & - & - \\ 
	LES & 11.2 & 21.5 & 27.6 & 9.46 & 15.3 & 18.6 \\ 
	SAS & 13.0 & 22.6 & 28.7 & 10.1 & 15.1 & 18.2 \\ 
	LES vs DNS & 1.24\% & 1.10\% & 2.75\% & - & - & - \\ 
	SAS vs LES & 16.7\% & 4.99\% & 4.08\% & 7.25\% & -1.45\% & -2.03\% \\ 
		\bottomrule
	\end{tabular}
	\label{tab:benchmarking}
\end{table}

It can be seen that LES and DNS results agree very well with only a minor discrepancy near the channel wall for $\textit{Pr}=15$, potentially due to the need for increased mesh refinement when simulating high-$\textit{Pr}$ fluids mentioned in the previous section. The SAS results show some disagreement, especially at low Prandtl numbers. It also appears that increasing the Reynolds number improves agreement. The dependence of the error on the Prandtl number could be explained by the fact that, as discussed in \cite{DiMarcello2010}, for low $\textit{Pr}$, the dependence of the SAS result on the modelling of $\textit{Pr}_t$ increases. In other words, the model error can be expected to increase for low $\textit{Pr}$. Similarly, the SAS was developed to analyse turbulent flow, while at Reynolds numbers $\textit{Re}_\delta$ close to $\sim$\num{2500}, the flow is in fact approaching the quasi-turbulent regime, for which model corrections are typically necessary \cite{Meyer2019}. Nevertheless, the agreement is reasonable overall.

Based on the presented results, we consider our LES approach to be successfully benchmarked against the reference high-fidelity DNS solution. Additionally, the performance of the SAS method is considered acceptable for broader parametric exploration performed in the next section.

\section{Reduced-order parametric exploration}
\label{sec:rom}
\begin{figure}
	\centering
	\includegraphics[width=\linewidth]{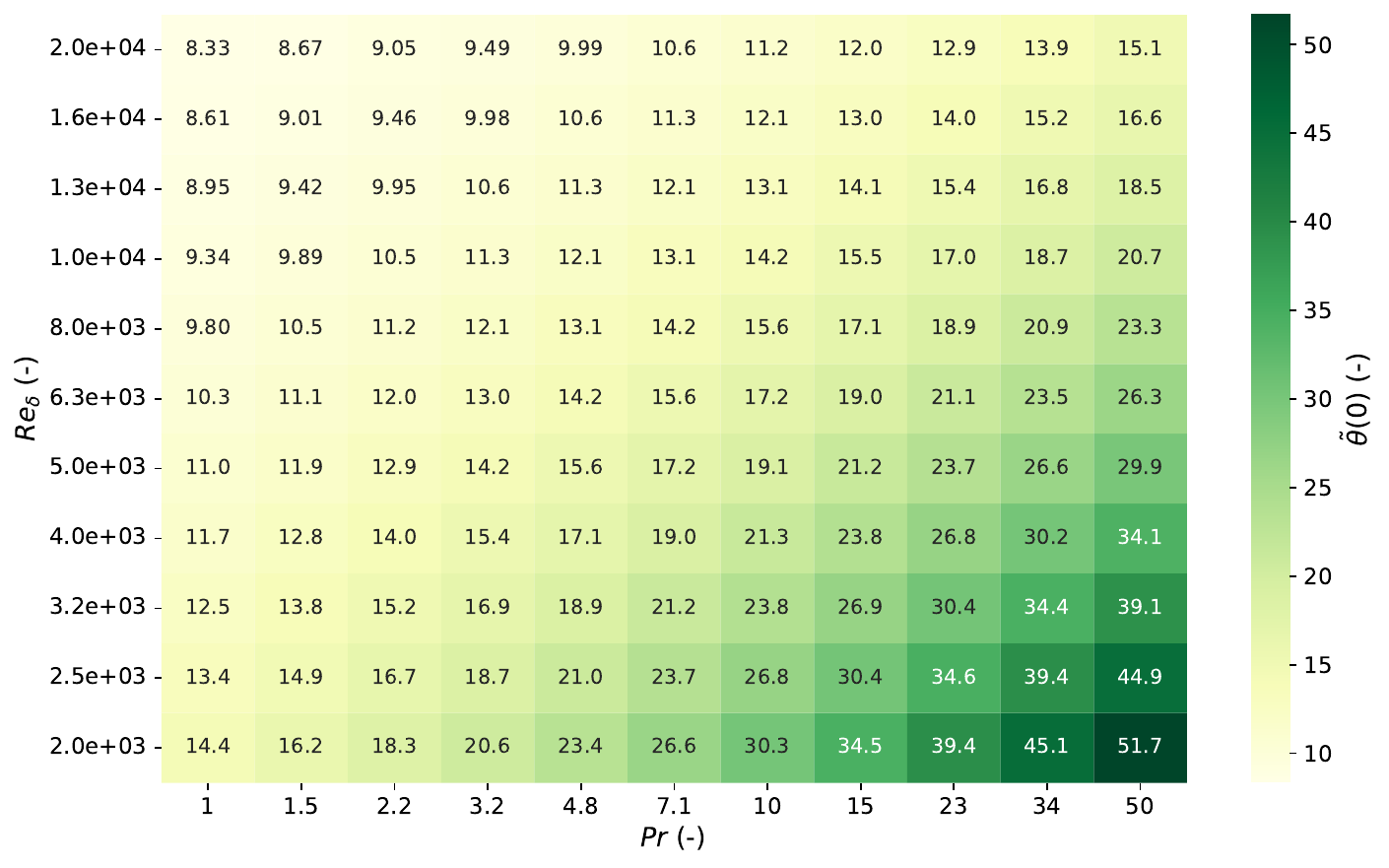}
	\caption{Non-dimensional temperature (Eq.\ \ref{eq:thetatilde}) at the wall evaluated as a function of Reynolds and Prandtl numbers using the SAS approach. Note that both axes are logarithmic.}
	\label{fig:devgridj}
\end{figure}

In the previous section, we ascertained the acceptable performance of the SAS method for analysing the effect of internal heating on the channel temperature distribution. Here we used the SAS approach to evaluate this effect for a broad range of Reynolds and Prandtl numbers. As our figure of merit, we selected the non-dimensional temperature (Eq.\ \ref{eq:thetatilde}) at the wall of the channel with the results shown in Fig.\ \ref{fig:devgridj}. The lower limit on $\textit{Pr}$, i.e.\ $\textit{Pr}_\textit{\textit{min}}=1$ was chosen based on the conclusions of the discussion in the previous section, while the lower limit on $\textit{Re}_\delta$, $\textit{Re}_{\delta,\textit{min}}=2000$, was taken as twice the Reynolds number for transition to quasi-turbulent flow for parallel-plate channel flows \cite{Sano2016}. The upper limits for these parameters were chosen somewhat arbitrarily to bracket the parameter range useful for the analysis of MSRs.

The results show that increasing Prandtl number and decreasing Reynolds number both increase the wall deviation from bulk temperature. This is not surprising, as increasing Reynolds number promotes turbulence and, therefore, smooths out thermal gradients. Conversely, increasing Prandtl number makes thermal gradients more pronounced and, thus, acts to increase the non-uniformity of temperature induced by internal heating. The data from Fig.\ \ref{fig:devgridj} were fitted by a power-law-type correlation of the form:
\begin{equation}
	\tilde{\theta}^{\textit{fit}}(0) = 1940 \cdot \textit{Re}_\delta^{-0.710} \cdot \textit{Pr}^{0.420} + 6.58.
	\label{eq:thetafit}
\end{equation}
For this fit, $R^2=0.9994$, the mean absolute error for the data in Fig.\ \ref{fig:devgridj} is 1.0\%, and the highest absolute error of 6.2\% is localised in the low $\textit{Re}_\delta$ and low $\textit{Pr}$ region. Equation \ref{eq:thetafit} shows that the effect of Reynolds number is stronger than the effect of Prandtl number. We also note in passing that, for developed laminar flow, Eq.\ \ref{eq:tw_lam} predicts for $j=0$:
\begin{equation}
	\tilde{\theta}(0)  = \frac{3}{35}\cdot\textit{Re}_\delta\cdot \textit{Pr},
\end{equation}
i.e.\ both $\textit{Re}_\delta$ and $\textit{Pr}$ tend to increase the non-dimensional wall temperature deviation. 

\begin{figure}
	\centering
	\includegraphics[width=\linewidth]{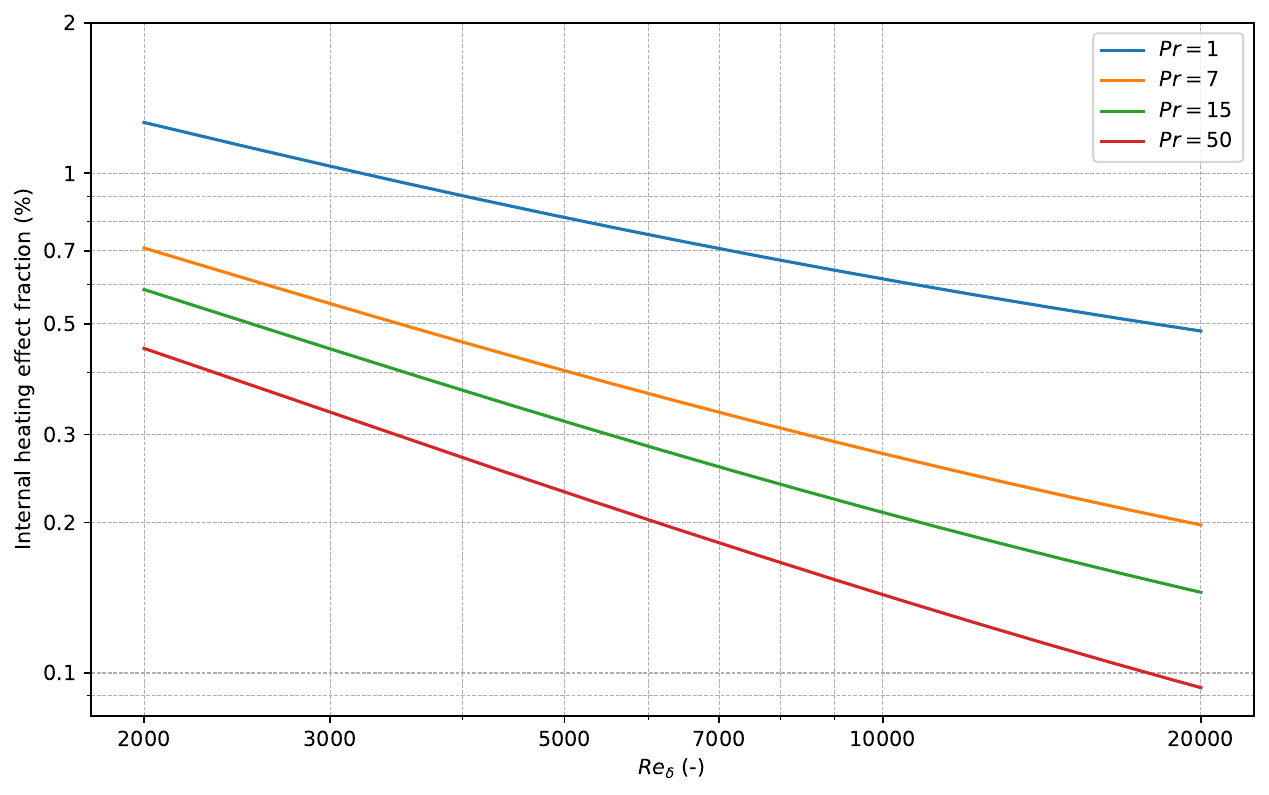}
	\caption{Fractional effect of internal heat generation on wall temperature for equivalent wall and internal heating, Eq.\ \ref{eq:deltaTrat}, in percent, as a function of Reynolds number for selected values of Prandtl numbers. Note that both axes are logarithmic.}
	\label{fig:ratio}
\end{figure}

We now consider the relation between the wall temperature deviation and Nusselt number due to internal heating based on hydraulic diameter $D_h = 4 \delta$ (Eq.\ \ref{eq:tw}):
\begin{equation}
	\textit{Nu}_q = 16\cdot \textit{Re}_\delta\cdot \textit{Pr}\ /\ \tilde{\theta}(0),
\end{equation}
and apply the fitted correlation:
\begin{equation}
	\textit{Nu}_q = \frac{16\cdot\textit{Re}_\delta\cdot \textit{Pr}}{1940 \cdot \textit{Re}_\delta^{-0.710} \cdot \textit{Pr}^{0.420} + 6.58}.
\end{equation}
Assuming formally equivalent wall heating and internal heat generation, i.e.\ $j=q\delta$, the fractional effect on wall temperature can be compared as:
\begin{equation}
	\zeta = \frac{\Delta T_{w,q}}{\Delta T_w} = \frac{\Delta T_{w,q}}{\Delta T_{w,j}+\Delta T_{w,q}} = \frac{1/\textit{Nu}_q}{1/\textit{Nu}_j + 1/\textit{Nu}_q}.
	\label{eq:deltaTrat}
\end{equation}
For $\textit{Nu}_j$ in the turbulent and quasi-turbulent regimes, a representative correlation can be chosen for example as follows (written here under the assumption of constant material properties) \cite{Meyer2019}:
\begin{equation}
	\textit{Nu}_j = 0.018 \cdot (4\textit{Re}_\delta)^{-0.25} \cdot(4\textit{Re}_\delta-500)^{1.07} \cdot \textit{Pr}^{0.42}.
\end{equation}
Under the assumption of this model, Fig.\ \ref{fig:ratio} shows the evaluation of Eq.\ \ref{eq:deltaTrat} for selected Prandtl numbers over the considered range of Reynolds numbers in percent. Clearly, internal heat generation has a significantly lower impact on wall temperature, and, by extension, on the channel temperature profile than wall heating of equivalent magnitude and the effect decreases with both Reynolds number and Prandtl number. We remark that using Eqs.\ \ref{eq:nuj_laminar} and \ref{eq:nuq_laminar}, the significance of internal heating is found to be low even for laminar flow with $\zeta_{\textit{laminar}} = 4.22$\%.

\begin{figure}
	\centering
	\includegraphics[width=\linewidth]{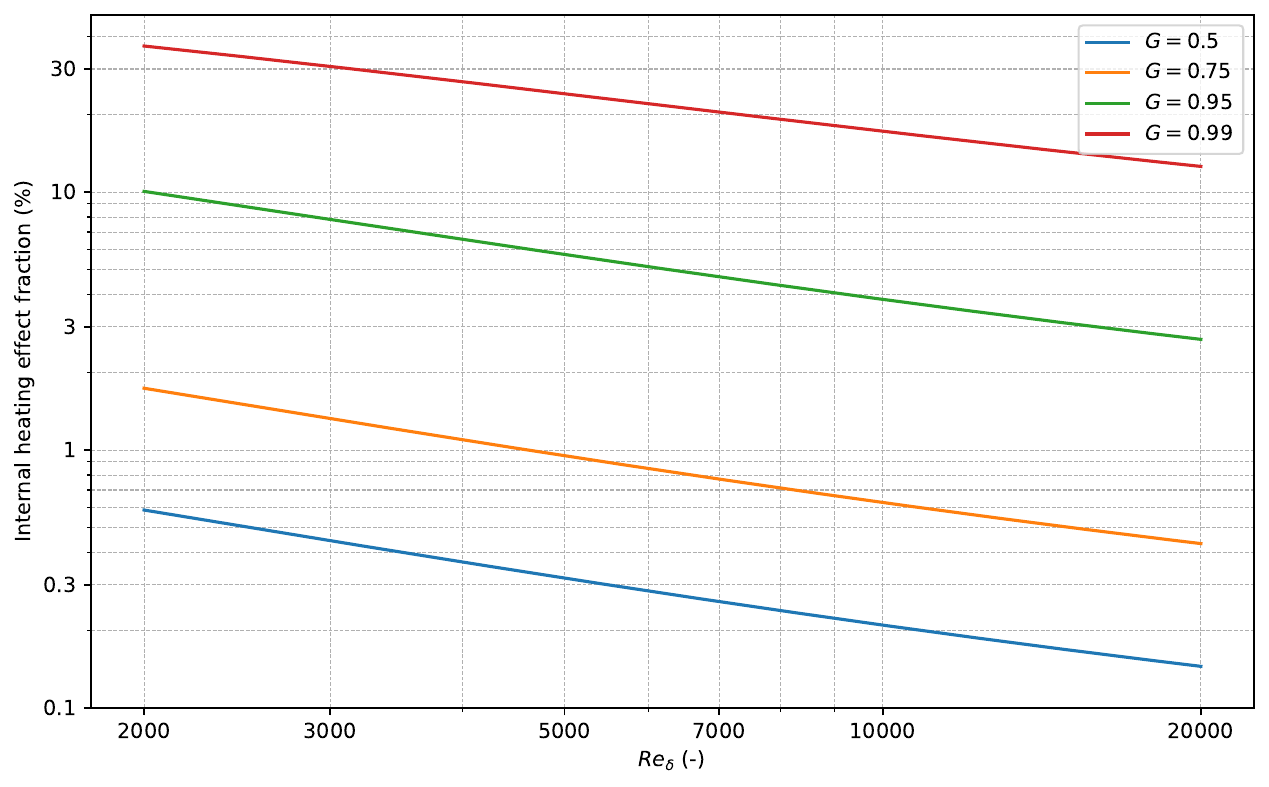}
	\caption{Fractional effect of internal heat generation on wall temperature, Eq.\ \ref{eq:deltaTrat_g}, in percent, as a function of Reynolds number for selected values of internal heating fraction $G$, Eq.\ \ref{eq:g}. Prandtl number is fixed as $15$. Note that both axes are logarithmic.}
	\label{fig:ratiog}
\end{figure}

It is, however, important to remember that in the reactor core of a thermal MSR only about 7\% of heat is typically deposited in the solid structures such as graphite \cite{Haubenreich1964} with the remainder being generated directly in the liquid fuel. Therefore, in a practical configuration, the previous analysis must be modified considering the relative importance of solid and liquid heat deposition, represented by the non-dimensional number $G$ (Eq.\ \ref{eq:g}). Subsequently, Eq.\ \ref{eq:deltaTrat} can be generalised as follows:
\begin{equation}
	\zeta = \frac{G/\textit{Nu}_q}{(1-G)/\textit{Nu}_j + G/\textit{Nu}_q}.
	\label{eq:deltaTrat_g}
\end{equation} 
Considering then the representative value of $\textit{Pr}$ for molten salts, $\textit{Pr}=15$, Fig.\ \ref{fig:ratiog} shows the effect of internal heat generation on wall temperature for selected values of $G$. Note that the case $G=0.5$ corresponds to the data in Fig.\ \ref{fig:ratio}. Evidently, the effect of internal heating becomes significant only for values of $G$ approaching 1. For completeness, for developed laminar flow, Eq.\ \ref{eq:deltaTrat_g} simplifies into:
\begin{equation}
	\zeta_{\textit{laminar}} = \frac{G}{68(1-G)/3 + G},
\end{equation} 
which evaluates to 36.9\% for 7\% wall heating, i.e. $G=0.93$. For turbulent flow, $\zeta$ would remain below 10\% in this situation, as can be discerned from Fig.\ \ref{fig:ratiog}. Overall, our analysis suggests that the role of internal heating on the temperature field in the channel is generally minor, especially for turbulent flow. When considering a concrete configuration, the framework presented here can be adapted with relative ease and a direct quantitative assessment of the importance of internal heat generation can then be made.

\section{Conclusions}
\label{sec:conc}
In this work, we focussed on an identified gap in understanding the effects of internal heating on temperature profiles in parallel-plate channel geometries, with a focus on turbulent flow configurations. This phenomenon is considered to be especially relevant to the design of liquid-fuelled nuclear reactor systems. For our investigations, we used three thermal-hydraulic approaches of varying degrees of fidelity, ranging from DNS through LES to a semi-analytical solution tool. We first introduced the considered problem, i.e.\ periodic developed flow in an infinite parallel-plate geometry, as well as the simulation tools. Subsequently, we selected a limited number of cases to benchmark the tools against one another. Very good agreement between DNS and LES results was found and the semi-analytical approach agreed satisfactorily with the LES reference.  To the best of the authors' knowledge, this work represents the first attempt for rigorous benchmarking of simulation methods for internally-heated fluids using DNS as a high-fidelity reference.

After we gained confidence in the performance of the semi-analytical approach, we used it for a reduced-order parametric exploration of the problem. It was found that, for turbulent flow, the effect of internal heating on the temperature profile becomes more significant with increasing Prandtl number and decreasing Reynolds number in the considered parameter range. The effect of Reynolds number was found to be the more important one of the two. Then, the effect of internal heating was compared to the effect of heating through the channel wall. The results of this comparison showed that internal heating is typically much less important than wall heating, especially in turbulent flow configuration. Nevertheless, we highlighted the fact that in liquid-fuelled reactors, such as MSRs, most of nuclear heat is deposited directly in the fuel, i.e.\ internal heating is generally significantly stronger than wall heating. However, even with this consideration in mind, it was found that internal heating becomes significant for shaping the temperature field in the fluid only when it represents an overwhelming share of the total heat deposition in the system.

In future work, we will extend our analysis by investigating the effects and configurations omitted in this paper. This includes but is not limited to: finite channel length considerations, role of non-uniform heating and variable material properties, explicit simulation of conjugate heat transfer, investigation of other flow regimes such as natural circulation or transitional turbulence, and transient analysis. Still further, we will explore the importance and characteristics of multi-phase flow in molten salts with and without internal heat generation. Overall, we expect that our simulation work will lay a foundation for future experimental work needed to validate our numerical results.

% Optional: competing interest acknowledgements
%\section*{Declaration of competing interest}

%\section*{Acknowledgements}

%------ References ------
\clearpage
\printbibliography

\end{document}